\documentclass[letterpaper,english,reprint,superscriptaddress,longbibliography]{revtex4-1}
\usepackage{lmodern}

\usepackage[T1]{fontenc}
\usepackage[utf8]{inputenc}
\usepackage{xcolor}
\usepackage{pdfcolmk}
\usepackage{babel}
\usepackage{float}
\usepackage{amsmath}
\usepackage{amssymb}
\usepackage{graphicx}
\PassOptionsToPackage{normalem}{ulem}
\usepackage{ulem}
\usepackage[unicode=true,pdfusetitle,
 bookmarks=false,
 breaklinks=false,pdfborder={0 0 0},pdfborderstyle={},backref=false,colorlinks=false]
 {hyperref}

\makeatletter

\pdfpageheight\paperheight
\pdfpagewidth\paperwidth

\providecolor{lyxadded}{rgb}{0,0,1}
\providecolor{lyxdeleted}{rgb}{1,0,0}

\DeclareRobustCommand{\lyxsout}[1]{\ifx\\#1\else\sout{#1}\fi}

\makeatother

\begin{document}
\title{Extended polarized semiclassical model for quantum-dot cavity QED
and its application to single-photon sources}
\author{H.J. Snijders}
\affiliation{Huygens-Kamerlingh Onnes Laboratory, Leiden University, P.O. Box 9504,
2300 RA Leiden, The Netherlands}
\author{D.N.L. Kok}
\affiliation{Huygens-Kamerlingh Onnes Laboratory, Leiden University, P.O. Box 9504,
2300 RA Leiden, The Netherlands}
\author{M.F. van de Stolpe}
\affiliation{Huygens-Kamerlingh Onnes Laboratory, Leiden University, P.O. Box 9504,
2300 RA Leiden, The Netherlands}
\author{J. A. Frey}
\affiliation{Department of Physics, University of California, Santa Barbara, California
93106, USA}
\author{J. Norman}
\affiliation{Department of Electrical \& Computer Engineering, University of California,
Santa Barbara, California 93106, USA}
\author{A. C. Gossard}
\affiliation{Department of Electrical \& Computer Engineering, University of California,
Santa Barbara, California 93106, USA}
\author{J. E. Bowers}
\affiliation{Department of Electrical \& Computer Engineering, University of California,
Santa Barbara, California 93106, USA}
\author{M.P. van Exter}
\affiliation{Huygens-Kamerlingh Onnes Laboratory, Leiden University, P.O. Box 9504,
2300 RA Leiden, The Netherlands}
\author{D. Bouwmeester}
\affiliation{Huygens-Kamerlingh Onnes Laboratory, Leiden University, P.O. Box 9504,
2300 RA Leiden, The Netherlands}
\affiliation{Department of Physics, University of California, Santa Barbara, California
93106, USA}
\author{W. Löffler}
\email{loeffler@physics.leidenuniv.nl}

\affiliation{Huygens-Kamerlingh Onnes Laboratory, Leiden University, P.O. Box 9504,
2300 RA Leiden, The Netherlands}
\begin{abstract}
We present a simple extension of the semi-classical model for a two-level
system in a cavity, in order to incorporate multiple polarized transitions,
such as those appearing in neutral and charged quantum dots (QDs),
and two nondegenerate linearly polarized cavity modes. We verify the
model by exact quantum master equation calculations, and experimentally
using a neutral QD in a polarization non-degenerate micro-cavity,
in both cases we observe excellent agreement. Finally, the usefulness
of this approach is demonstrated by optimizing a single-photon source
based on polarization postselection, where we find an increase in
the brightness for optimal polarization conditions as predicted by
the model.
\end{abstract}
\maketitle

\section{introduction}

Understanding the interaction of a two-level system, such as atomic
transitions or excitonic transitions in a semiconductor quantum dot
(QD), with an optical cavity mode, is key for designing efficient
single photon sources \citep{grange2015,somaschi_near-optimal_2016,ding_-demand_2016,wang2019g}
and photonic quantum gates \citep{imamoglu_strongly_1997} for quantum
networks \citep{kimble_quantum_2008}. Traditionally, the interaction
of a two-level quantum system with an electromagnetic mode is described
by the Jaynes-Cummings model, which can be approximated in the so-called
semi-classical approach, where the light field is treated classically
and atom-field correlations are neglected. We focus here on QD-cavity
systems in the weak coupling ``bad cavity'' regime $\left(g\ll\kappa\right)$.
The transmission amplitude in the semi-classical approximation is
given by \citep{armen_low-lying_2006,waks_dipole_2006,auffeves-garnier_giant_2007,loo_optical_2012,majumdar2012c,hu_extended_2015,bakker_polarization_2015}

\begin{equation}
t=\eta_{out}\frac{1}{1-2i\Delta+\frac{2C}{1-i\Delta^{'}}}.\label{Ch2-eq-transmission-electric-field-FP}
\end{equation}
Here, $\eta_{out}$ is the probability amplitude that a photon leaves
the cavity through one of the mirrors, we assume two identical mirrors.
$\Delta=\left(f-f_{c}\right)/\kappa$ is the normalized detuning of
the laser frequency \footnote{All frequencies and rates here are ordinary and measured in hertz.}
$f$ with respect to the cavity resonance frequency $f_{c}$ and cavity
loss rate $\kappa$, $\Delta'=\left(f-f^{'}\right)/\gamma_{\perp}$
is the normalized detuning with respect to the QD resonance frequency
$f^{'}$ and dephasing rate $\gamma_{\perp}=\frac{\gamma_{||}}{2}+\gamma^{*}$.
$\Delta$ is related to the round trip phase $\phi$ by $\phi\approx\frac{2\pi\Delta}{\mathcal{{F}}}$
for small detuning $\Delta$, and $\mathcal{{F}}$ is the finesse
of the cavity. The coupling of the QD to the cavity mode is given
by the cooperativity parameter $C=\frac{g^{2}}{\kappa\gamma_{\perp}}$,
where the QD-cavity coupling strength is $g$. In Appendix A, we show
how Eq.~\ref{Ch2-eq-transmission-electric-field-FP} can be derived
in a fully classical way. The main limitation of semi-classical models
is that the population of the excited state is not taken into account,
as well as phonon-assisted transitions, spin flips, and other interactions
with the environment. 

In this paper, Eq.~(\ref{Ch2-eq-transmission-electric-field-FP})
is extended to take into account two orthogonal linearly-polarized
fundamental optical cavity modes, and multiple polarized QD transitions
oriented at an arbitrary angle relative to the cavity polarization
axes. This extension is important because it is experimentally very
challenging to produce perfectly polarization degenerate micro-cavities
\citep{bonato_tuning_2009,frey_electro-optic_2018}, and the slightly
non-polarization degenerate case has attracted attention recently
\citep{anton_tomography_2017,he_polarized_2018,wang2019g}. It is
essential to have access to a good analytic model, for instance to
numerically fit experimental data to derive the system parameters,
or to optimize the performance of a single-photon source; this is
very time-consuming using exact quantum master equation simulations.
Exemplary code of our model is available online \citep{snijders2018f}.
We compare our model to experimental data as well as numerical solutions
of the quantum master equation, and we demonstrate that it can be
used to significantly increase the brightness of a single-photon source.
We focus here on Fabry-Perot type QD-cavity systems but our results
are valid for a large range of cavity QED systems.

\section{Extended semi-classical model}

\begin{figure}
\begin{centering}
\includegraphics[width=1\columnwidth]{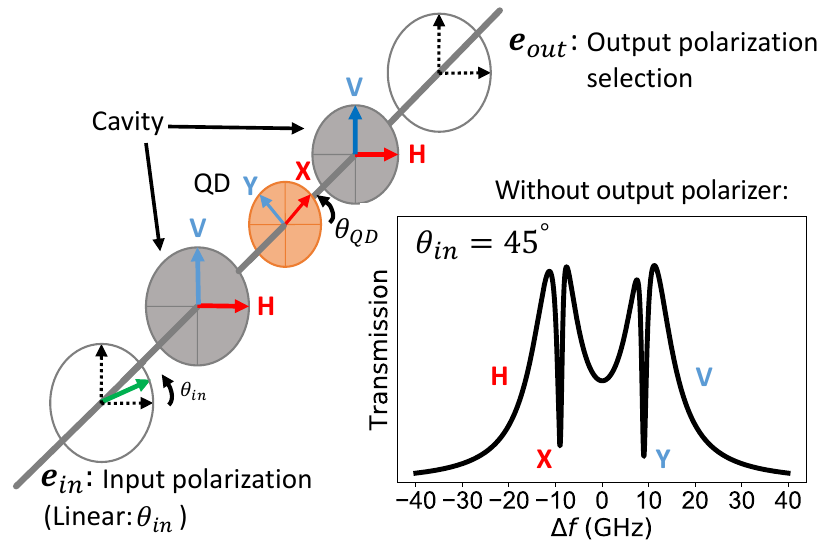}
\par\end{centering}
\caption{Sketch of a polarized cavity–neutral QD system. H and V denote the
linearly polarized cavity modes, X and Y represent the dipole polarization
axes of the QD at an angle $\theta_{QD}$ with respect to the H cavity
polarization, and $\mathbf{e}_{in}$ and $\mathbf{e}_{out}$ indicate
the incident polarization and output polarization postselection. The
inset shows the transmission spectrum calculated with the extended
semiclassical model for incident linear polarized light ($\theta_{in}=45^{\circ}$).
The difference in dip depth between the X and Y transitions is due
to the specific QD dipole orientations ($\theta_{QD}$). Here, no
polarization postselection is done. The parameters are $f_{H}=-10\,$GHz,
$f_{V}=10\,$GHz, $f_{X}^{'}=-9\,$GHz, $f_{Y}^{'}=9\,$GHz, $\theta_{QD}=10^{\circ}$.
\label{CH2-Fig1:SketchpolCQED}}
\end{figure}

To start the analysis, we show in Fig.~\ref{CH2-Fig1:SketchpolCQED}
a sketch of a polarized QD-cavity system with two cavity modes (H,V)
and two QD dipole transitions (X,Y). In order to demonstrate the complexity
of the transmission spectrum that appears is this case, we show in
the inset of Fig.~\ref{CH2-Fig1:SketchpolCQED} the transmission
of linearly polarized input light $\left(\theta_{in}=45^{\circ}\right)$
as a function of the relative laser frequency $\Delta f$. 

We now show how Eq. \ref{Ch2-eq-transmission-electric-field-FP} can
directly be extended to take care of all polarization effects, by
replacing the scalar quantities by appropriate Jones vectors and matrices.
To motivate the precise form, we first write Eq. \ref{Ch2-eq-transmission-electric-field-FP}
as its Taylor expansion

\[
\begin{array}{c}
t=\eta_{out}\left[1+\left(-2i\Delta+\frac{2C}{1-i\Delta^{'}}\right)+\right.\\
\left.\left(-2i\Delta+\frac{2C}{1-i\Delta^{'}}\right)^{2}+\cdots\right],
\end{array}
\]

where we now can clearly identify contributions from the cavity and
from the QD. This form reminds us of the multiple roundtrips happening
in a Fabry-Perot cavity, we show a complete derivation of Eq. \ref{Ch2-eq-transmission-electric-field-FP}
in the Appendix. In the polarization basis of the cavity, the normalized
detuning phase $2i\Delta$ becomes the Jones matrix
\begin{equation}
\left(\begin{array}{cc}
2i\Delta_{H} & 0\\
0 & 2i\Delta_{V}
\end{array}\right),\label{CH2-eq-matrixpolcav}
\end{equation}

where $\Delta_{m}=\left(f-f_{m}\right)/\kappa_{m}$ for $m=H,V$ are
the normalized laser detunings from the polarized cavity resonances
at frequencies $f_{m}$. The interaction with the QD modifies the
round-trip phase, but because of a possible misalignment of the dipole
axes of the QD transitions and the cavity polarization basis, we have
to calculate the QD effect in its own basis, which is accomplished
by $R_{-\theta_{QD}}XR_{\theta_{QD}}$, where $R_{\theta_{QD}}$ is
the 2D rotation matrix and $\theta_{QD}$ the rotation angle between
the cavity and quantum dot frame, see Fig.~\ref{CH2-Fig1:SketchpolCQED}.
There are many different transitions possible in QDs \citep{bayer_fine_2002,warburton_single_2013}.
This can be described by a transmission matrix $X$ composed of the
appropriate Jones matrices $J_{n}$ (see Table 2.1 in \citep{fowles_introduction_1989})
and the Lorentzian frequency-dependent phase shifts $\phi_{n}$:

\begin{equation}
X=\sum_{n}J_{n}\varphi_{n}=\sum_{n}J_{n}\frac{2C_{n}}{1-i\Delta_{n}^{'}},\label{X-matrix}
\end{equation}
$\Delta_{n}^{'}=\left(f-f_{n}^{'}\right)/\gamma_{\perp n}$ are the
normalized laser detunings from the QD resonances at $f_{n}^{'}$,
and $C_{n}$ are their cooperativity parameters. The case discussed
here is that of a neutral QD exciton, where $X=\varphi_{H}H+\varphi_{V}V$
which is equal to Eq. \ref{CH2-eq-matrixpolcav}. Note that, due to
the nature of semi-classical models, nonlinear (such as electromagnetically
induced transparency, EIT) and non-resonant effects (such as spin
relexation and phonon interactions) are not reproduced. The resulting
polarized Taylor-expanded expression for the transmission of the QD-cavity
system is then given by

\[
\begin{array}{c}
t_{tot}=\eta_{out}\left[I_{2\times2}+\left(-\left(\begin{array}{cc}
2i\Delta_{H} & 0\\
0 & 2i\Delta_{V}
\end{array}\right)+R_{-\theta_{QD}}XR_{\theta_{QD}}\right)+\right.\\
\left.\left(-\left(\begin{array}{cc}
2i\Delta_{H} & 0\\
0 & 2i\Delta_{V}
\end{array}\right)+R_{-\theta_{QD}}XR_{\theta_{QD}}\right)^{2}+\cdots\right].
\end{array}.
\]

Finally, we perform the reverse Taylor expansion, and obtain the full
transmission amplitude matrix, which is the main result of this paper:

\begin{equation}
t_{tot}=\eta_{out}\left[I_{2\times2}-\left(\begin{array}{cc}
2i\Delta_{H} & 0\\
0 & 2i\Delta_{V}
\end{array}\right)+R_{-\theta_{QD}}XR_{\theta_{QD}}\right]^{-1}.\label{CH2-eq-polmodel-4}
\end{equation}

Note that this result could have been directly obtained by plugging
in the appropriate matrix expressions into Eq. \ref{Ch2-eq-transmission-electric-field-FP}.
Experimentally most relevant is the scalar transmission amplitude
for the case that the cavity-QED system is placed between an input
and output polarizer. This can be obtained by $\mathbf{e}_{out}^{T}t_{tot}\mathbf{e}_{in}$,
where $\mathbf{e}_{in}=\left(E_{in}^{H},E_{in}^{V}\right)^{T}$ and
$\mathbf{e}_{out}=\left(E_{out}^{H},E_{out}^{V}\right)^{T}$ are the
input and output Jones vectors or polarizations, also shown in the
published code examples \citep{snijders2018f}.

\begin{figure}[H]
\begin{centering}
\includegraphics[width=1\columnwidth]{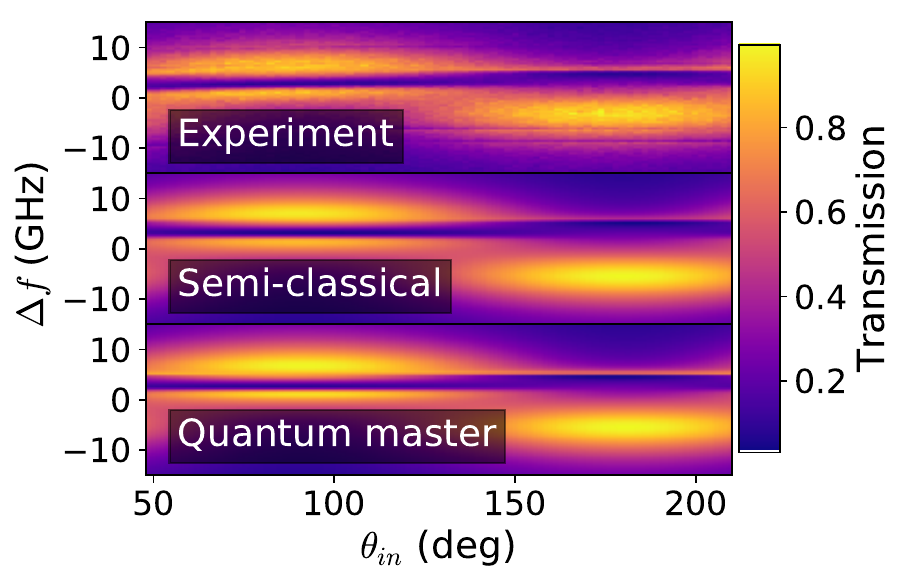}
\par\end{centering}
\caption{False color plots of the cavity transmission as function of laser
frequency and linear input polarization orientation. The three false
color plots show the experimental data, corrected for reduced detection
efficiency, polarized semi-classical theory results based on Eq.~(\ref{CH2-eq-polmodel-4})
and numerical simulations based on the quantum master model. \label{Ch2-Fig2:False-color-plotsQD-1}}
\end{figure}

We now compare our model to experiments and exact numerical simulations
of the quantum master equation using QuTiP \citep{johansson_qutip:_2012,johansson_qutip_2013},
for a neutral QD in a polarization non-degenerate cavity. The device
consists of a micropillar cavity with an embedded self-assembled QD
\citep{snijders_purification_2016}. In Fig.~\ref{Ch2-Fig2:False-color-plotsQD-1},
a false color plot of the measured transmission as a function of the
relative laser detuning and the orientation of linearly polarized
input laser is shown. By careful fitting of our model to the experimental
data we obtain excellent agreement (see Fig.~\ref{Ch2-Fig2:False-color-plotsQD-1})
using the following parameters: $\theta_{QD}=94^{\circ}\pm2^{\circ}$,
cavity splitting $f_{V}-f_{H}=10\pm0.1$ GHz, QD fine-structure splitting
$f_{Y}^{'}-f_{X}^{'}=2\pm0.1$ GHz, $\kappa=11.1\pm0.1$ GHz, $g=1.59\pm0.08$
GHz and $\gamma_{\parallel}=0.32\pm0.15$ GHz ($\gamma^{*}$ set to
zero). From this, we obtain for both transitions the cooperativity
$C=1.42\pm0.5.$ Inserting these parameters in the quantum master
equation for this system \citep{snijders_purification_2016} we again
find excellent agreement (see Fig.~\ref{Ch2-Fig2:False-color-plotsQD-1}).
In Appendix C we show that, for low mean photon number, the numerical
results from the quantum master equation are equal to our extended
semi-classical approach. 

\section{Application to single photon sources}

Now we show that our model can be used to optimize the polarization
configuration for quantum-dot based single-photon sources, in particular
the single-photon purity (determined by the second-order correlation
$g^{2}(0)$), and the brightness. To calculate $g^{2}(0)$, we need
to take into account two contributions: First, single-photon light
that has interacted with the QD, $\rho^{sp}(x)=x\left|1\right\rangle \left\langle 1\right|+\left(1-x\right)\left|0\right\rangle \left\langle 0\right|$,
where $x$ is the mean photon number. Second, ``leaked'' coherent
laser light, $\rho^{coh}(\alpha)$, with the mean photon number, $\left\langle n^{coh}\right\rangle =\left|\alpha\right|^{2}$,
where $\left|\alpha\right|^{2}$ can be determined by tuning the QD
out of resonance. With a weighting parameter, $\xi$, the density
matrix of the total detected light can be written as
\begin{equation}
\rho^{tot}=\left[\xi\rho^{sp}(x)+(1-\xi)\rho^{coh}(\alpha)\right].\label{Ch2-eq-rho-density-1}
\end{equation}

After determining $\rho^{tot}$, it is straightforward to obtain $g^{(2)}(0)$
of the total transmitted light \citep{proux_measuring_2015}.

\begin{figure}
\begin{centering}
\includegraphics[width=1\columnwidth]{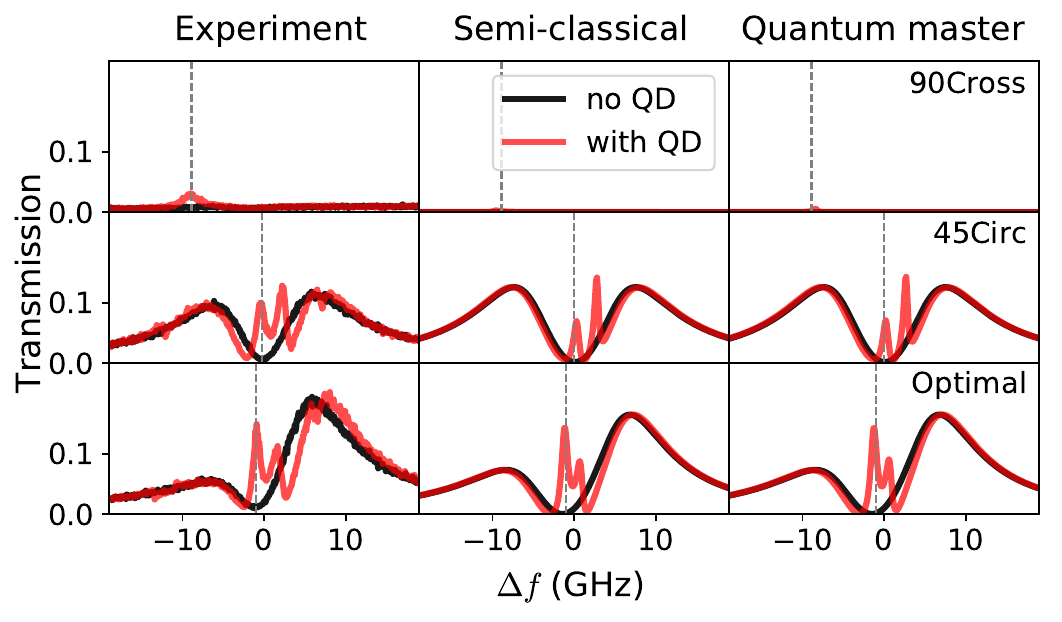}
\par\end{centering}
\caption{Measured (left), semi-classical simulated (middle) and quantum master
simulated(right) transmitted intensity as a function of the relative
laser frequency, with and without the QD, and for the three polarization
configurations 90Cross (top), 45Circ (center), and Optimal (bottom).
For constant laser power, the measured single-photon intensity (frequency
indicated by the dashed vertical line) of the optimal configuration
is about $3\times$ ($1.6\times$) higher compared to the 90Cross
(45Circ) configuration. \label{Ch2-Fig3:crosspolplots}}
\end{figure}

To find the optimal polarization condition, we numerically optimize
the input and output polarization, as well as the quantum dot and
laser frequency, in order to maximize the light that interacted with
the QD transition (single photon light), and to minimize the residual
laser light. This is easily feasible because calculation of the extended
semiclassical model is fast. We compare the optimal result to the
conventional polarization conditions 90Cross (excitation of the H-
and detection along the V-cavity mode) and ``45Circ''. For 45Circ,
the system is excited with $45^{\circ}$ linear polarized light and
we detect a single circular polarization component. This works because,
in this configuration, the birefringence of the cavity modes functions
as a quarter wave plate. Fig.~\ref{Ch2-Fig3:crosspolplots} compares
the theoretical prediction to the experimental data for these cases,
each with and without the QD. These results show almost perfect agreement
between experiment and theory. Only for the 90Cross configuration,
the experimental data is slightly higher than expected, which we attribute
to small changes of the polarization axes of the QD induced by the
necessary electrostatic tuning of the QD resonance.

The optimal polarization condition is found for the input polarization
Jones vector $\mathbf{e}_{in}=\left(\begin{array}{cc}
0.66, & 0.50-0.57i\end{array}\right)^{T}$ and output polarization $\mathbf{e}_{out}=\left(\begin{array}{cc}
0.66, & 0.50+0.57i\end{array}\right)^{T}$. For this case, the single photon intensity is about $3\times$ higher
compared to the 90Cross configuration. We emphasize that this optimal
configuration can hardly be found experimentally because the parameter
space, polarization conditions and QD and laser frequencies, is too
large. Instead, numerical optimization has to be done, for which a
simple analytical model, like the one presented here, is essential.
Again we compare our extended semi-classical model to exact numerical
simulations from the QuTiP to verify the validity of our model and
the experimental results (see Fig.~\ref{Ch2-Fig3:crosspolplots}).
Because here, the complex transmission amplitudes of both polarizations
interfere, we can conclude from the good agreement that not only the
transmission but also the transmission phases of Eq. \ref{CH2-eq-polmodel-4}
are correctly reproduced by the model.

For the configurations shown in Fig.~\ref{Ch2-Fig3:crosspolplots},
we now perform power-dependent continuous-wave measurements to determine
the experimental brightness and $g^{(2)}(0)$. The laser is locked
at the optimal frequency determined by the model (dashed vertical
line in Fig.~\ref{Ch2-Fig3:crosspolplots}), and the single photon
count rate, as well as the second-order correlation function, is measured
using a Hanbury-Brown Twiss setup. The photon count rate is the actual
count rate before the first lens, corrected for reduced detection
efficiency. Gaussian fits to $g^{(2)}(\tau)$ are used to determine
the second-order correlation function at zero time delay $g^{(2)}(0)$.

\begin{figure}
\begin{centering}
\includegraphics[width=1\columnwidth]{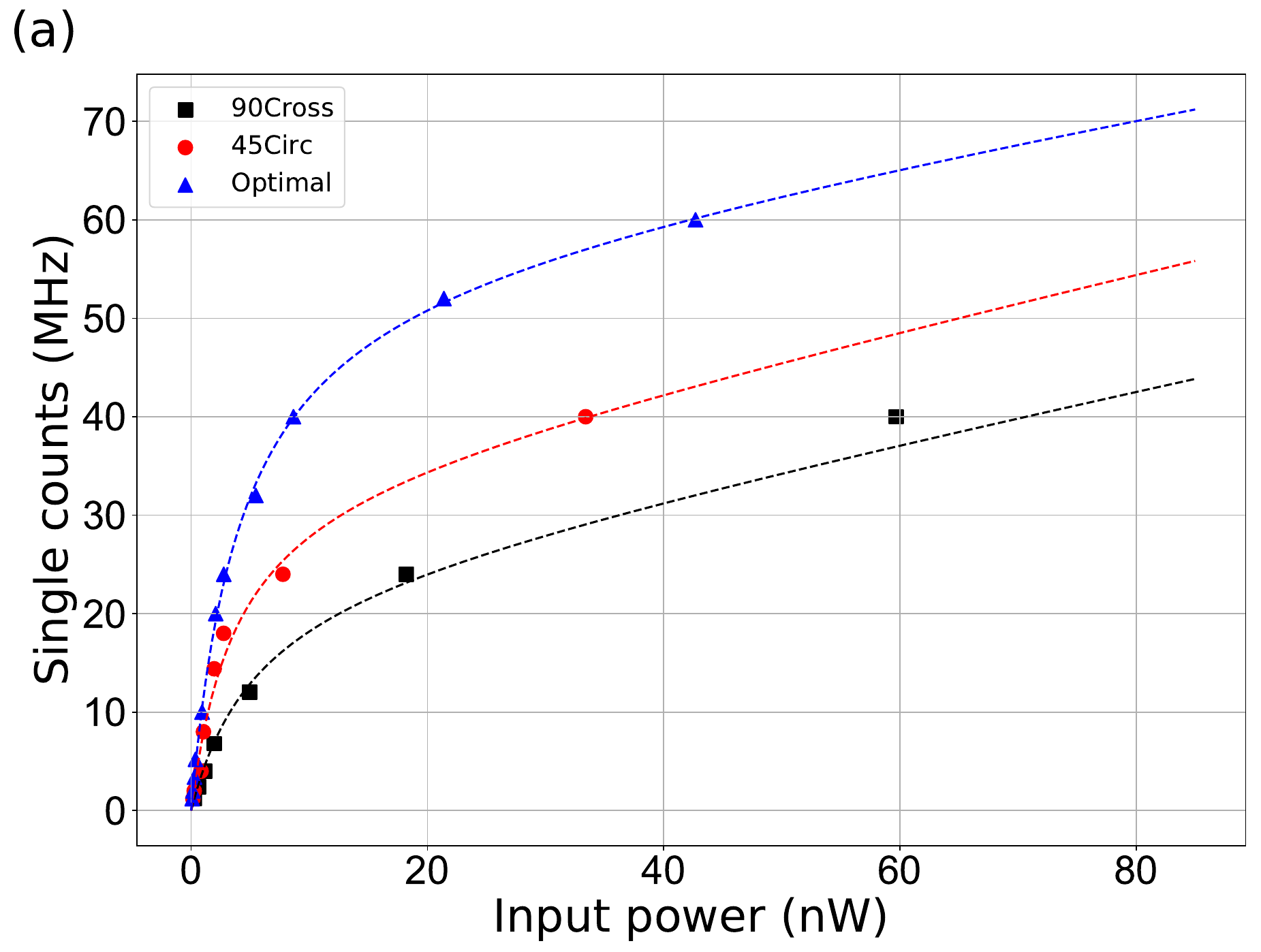}
\par\end{centering}
\begin{centering}
\includegraphics[width=1\columnwidth]{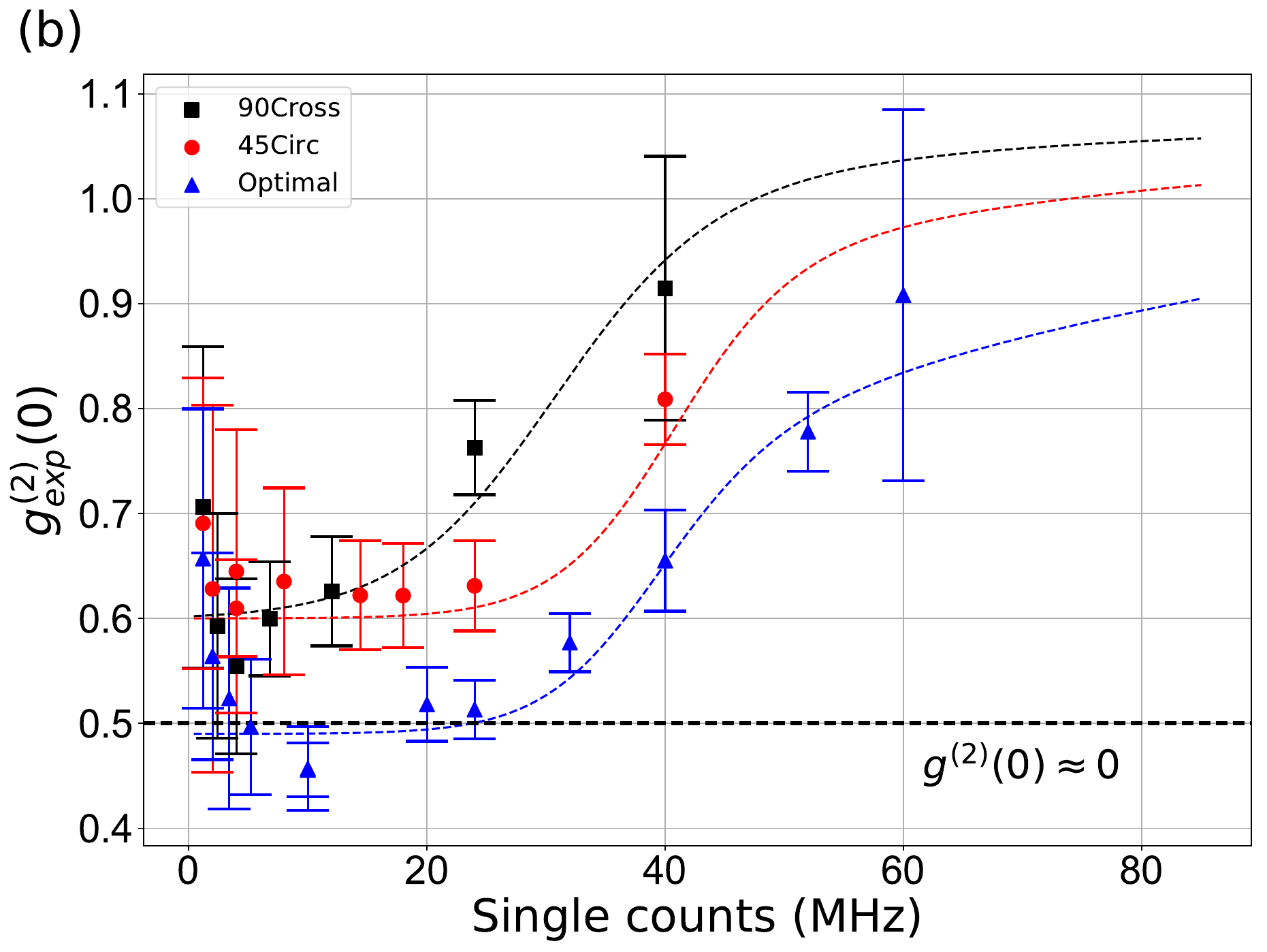}
\par\end{centering}
\caption{(a) Single-photon count rate \textbf{$\Gamma$} behind the first lens
as a function of the input laser power for the three polarization
configurations 90Cross (squares), 45Circ (circles), Optimal (triangles).
The dashed curves are fits to Eq.~(\ref{Ch2-eq-inputpower vs output counts})
and show good agreement. (b) $g_{exp}^{(2)}(0)$ as a function of
the measured single-photon count rate behind the first lens. The dashed
curves are the theoretical predictions as described in the text. The
increased size of the error bars at higher power is because the $g_{exp}^{(2)}(\tau)$
dip becomes small. \label{Ch2-Fig4:Singlephotonrate}}
\end{figure}

In Fig.~\ref{Ch2-Fig4:Singlephotonrate}(a), the single-photon count
rate is shown as a function of the input power, and in Fig.~\ref{Ch2-Fig4:Singlephotonrate}(b)
we show $g_{exp}^{(2)}(0)$ as a function of the single-photon count
rate. In Fig.~\ref{Ch2-Fig4:Singlephotonrate}(b), we see that, for
the optimal configuration, the single photon rate can be up to 24
MHz before the purity of the single-photon source decreases. This
means that, for the same purity, it is possible to increase the brightness
of the single-photon source by using different input and output polarization
configurations. Note that $g_{exp}^{2}(0)\approx0.5$ corresponds
to a real $g^{(2)}(0)\approx0$ due to detector jitter. The two-detector
jitter of $\approx500$ ps, which is of the same order as the the
cavity enhanced QD decay rate, explains the limited lower value of
$g_{exp}^{(2)}(\tau)$.

The data in Fig.~\ref{Ch2-Fig4:Singlephotonrate}(a) shows the interplay
between single-photon light scattered from the QD and leaked coherent
laser light. We observe a linear slope for high input power, which
corresponds to laser light that leaks through the output polarizer.
In Fig.~\ref{Ch2-Fig4:Singlephotonrate}(a) we fit the single photon
rate, $\Gamma$, using the formula \citep{snijders_fiber-coupled_2018}
\begin{equation}
\left(x+\left\langle n^{coh}\right\rangle \right)\gamma_{\perp}=\text{\ensuremath{\Gamma}}\frac{\frac{P}{P_{0}}}{1+\frac{P}{P_{0}}}+bP.\label{Ch2-eq-inputpower vs output counts}
\end{equation}

Here, \textbf{$b$ }is the fraction of leaked laser light, $P_{0}$
is the saturation power of the QD, and\textbf{ $\Gamma$} is the experimentally
obtained single photon rate of the QD. We find for the optimal condition
$P_{0}\approx3\,$nW, $\Gamma\approx40\,$MHz, and $b\approx0.5\,$MHz
nW$^{-1}$. This single photon rate is $25\,$\% of the maximal output
through one of the mirrors, based on the QD lifetime, $\gamma_{\perp}/2\approx160\,$MHz.
Calculating $g^{(2)}(0)$ using Eq.~(\ref{Ch2-eq-rho-density-1})
gives the predictions shown by the dashed curves in Fig.~\ref{Ch2-Fig4:Singlephotonrate}(b).
For these predictions, we use $\gamma_{\perp}=$ $320\,$MHz in order
to obtain the mean photon number $x$. Now, considering the detector
response, we estimate $\xi_{90}=0.05$ in Eq.~(\ref{Ch2-eq-rho-density-1})
for the 90Cross configuration, which allows us to derive $\xi_{45}=1.6\times\xi_{90}=0.10$
and $\xi_{opt}=3\times\xi_{90}=0.15$ using the data shown in Fig.
\ref{Ch2-Fig3:crosspolplots}. Here, $\xi$ corresponds to the single-photon
brightness as a result of the polarization projection. We see that
our theory is in good agreement with the experimental data in Fig.~\ref{Ch2-Fig4:Singlephotonrate}(b).

In principle, if the output polarizer could block all residual laser
light, a perfectly pure single-photon source is expected. In this
case, the brightness of the single-photon source is determined by
the polarization change that the QD-scattered single photons experience.
At high power, close to QD saturation, the QD also emits non-resonant
light, but its effect on the purity is limited in practice compared
to the effect of leaked laser light \citep{matthiesen_subnatural_2012}.

\section{conclusion}

In conclusion, we have proposed a polarized semi-classical cavity-QED
model, and confirmed its accuracy by comparison to experimental data
of a QD micro-cavity system. We have shown that this model enables
prediction and optimization of the brightness and purity of QD-based
single-photon sources, where we have obtained a $3\times$ higher
brightness compared to traditional cross-polarization conditions.
The model can also be used to optimize pulsed single-photon sources
by integrating over the broadened spectrum of the exciting laser.
\begin{acknowledgments}
We acknowledge funding from the European Union’s Horizon 2020 research
and innovation programme under grant agreement No. 862035 (QLUSTER),
from FOM-NWO (08QIP6-2), from NWO/OCW as part of the Frontiers of
Nanoscience program and the Quantum Software Consortium, and from
the National Science Foundation (NSF) (0901886, 0960331).
\end{acknowledgments}

\appendix

\section{Intuitive derivation of the semi-classical model}

Here we show an alternative, intuitive, derivation of Eq.~(1) in
the main text. We consider two equal mirrors with reflection coefficient
$r$ and transmission coefficient $t$ at a distance $L$, like a
Fabry-Pérot resonator. The round-trip phase $\phi_{0}$ in the electric
field propagation term, written in terms of the wavelength $\lambda_{0}$,
refractive index $n$ and length $L$ of the cavity, is:

\begin{equation}
\phi_{0}=\frac{2\pi}{\lambda_{0}}n\left(2L\right)=\frac{4\pi nL}{c}f,\label{Ch2-eq-FP-deriv-2}
\end{equation}
where $c$ is the speed of light and $f$ the frequency of the laser.
Since the laser frequency will be scanned across the resonance frequency
$f_{c}$ of the Fabry-Pérot cavity, it is convenient to write the
phase shift in terms of the relative frequency:

\begin{equation}
\phi=\frac{4\pi nL}{c}\left(f-f_{c}\right).\label{Ch2-eq-FP-deriv-3}
\end{equation}
Further, we assume that there is dispersion and loss in the cavity.
We quantify loss of the cavity by single pass amplitude loss $a_{0}$.
The QD transition is described by a harmonic oscillator. In the rotating
wave approximation, a driven damped harmonic oscillator has a frequency-dependent
response similar to a complex Lorentzian. Including cavity loss $a_{0}$,
QD loss $a_{QD}$ and Lorentzian dispersion, we obtain a field change
in half a round trip of

\begin{equation}
\exp\left(-a+i\frac{\phi}{2}\right),\;\;\;\;\text{where }a\equiv a_{0}+\frac{a_{QD}}{1-i\Delta^{'}}.\label{Ch2-eq-FP-deriv-4}
\end{equation}
Here, $\Delta^{'}=\left(f-f_{QD}\right)/\gamma_{\perp}$ with the
resonance frequency of the QD $f_{QD}$. By summing over all possible
round trips, the total transmission amplitude is

\begin{equation}
\begin{array}{c}
t_{tot}=tt\exp\left(-a+i\phi/2\right)\left[\sum_{n=0}^{\infty}\left(r^{2}\exp\left(-2a+i\phi\right)\right)^{n}\right]\end{array}\label{Ch2-eq-transmissionsum-1}
\end{equation}

which becomes

\begin{equation}
t_{tot}=\frac{t^{2}\exp\left(-a+i\phi/2\right)}{1-r^{2}\exp\left(-2a+i\phi\right)}.\label{Ch2-eq-transmissionsum-2}
\end{equation}
This formula can be written in a form similar to the semi-classical
model by considering $R\sim1$, small phase changes in the cavity
$\phi\ll1$, in combination with $a_{QD}\ll1$. This allows us to
use a Taylor expansion of the exponentials in Eq.~(\ref{Ch2-eq-transmissionsum-2}).
By including all first-order contributions and a few second-order
contributions, we write the complex transmission amplitude as

\begin{equation}
t_{tot}\approx\eta_{out}\frac{1}{1-2i\Delta+\frac{2C}{1-i\Delta^{'}}},\label{Ch2-eq-transmission-electric-field-FP-1}
\end{equation}
with the out-coupling efficiency 
\begin{equation}
\eta_{out}=\frac{1}{\sqrt{1+2a_{0}\left(\frac{1+R}{1-R}\right)}}.\label{Ch2-eq-fp-value-nout}
\end{equation}
In Appendix B, we show how to derive Eq.~(\ref{Ch2-eq-transmission-electric-field-FP-1})
and explain that the added higher order Taylor terms to write the
final formula in a compact form, are negligible. The out-coupling
efficiency $\eta_{out}$ gives the probability that a photon leaves
the cavity through one of the mirrors. In Eq.~(\ref{Ch2-eq-transmission-electric-field-FP-1}),
$\Delta$ is the normalized laser-cavity detuning and $\Delta'$ is
the normalized detuning with respect to the QD transition. 

\section{Detailed derivation of equation \ref{Ch2-eq-transmission-electric-field-FP-1}}

\begin{figure}[H]
\begin{centering}
\includegraphics[width=1\columnwidth]{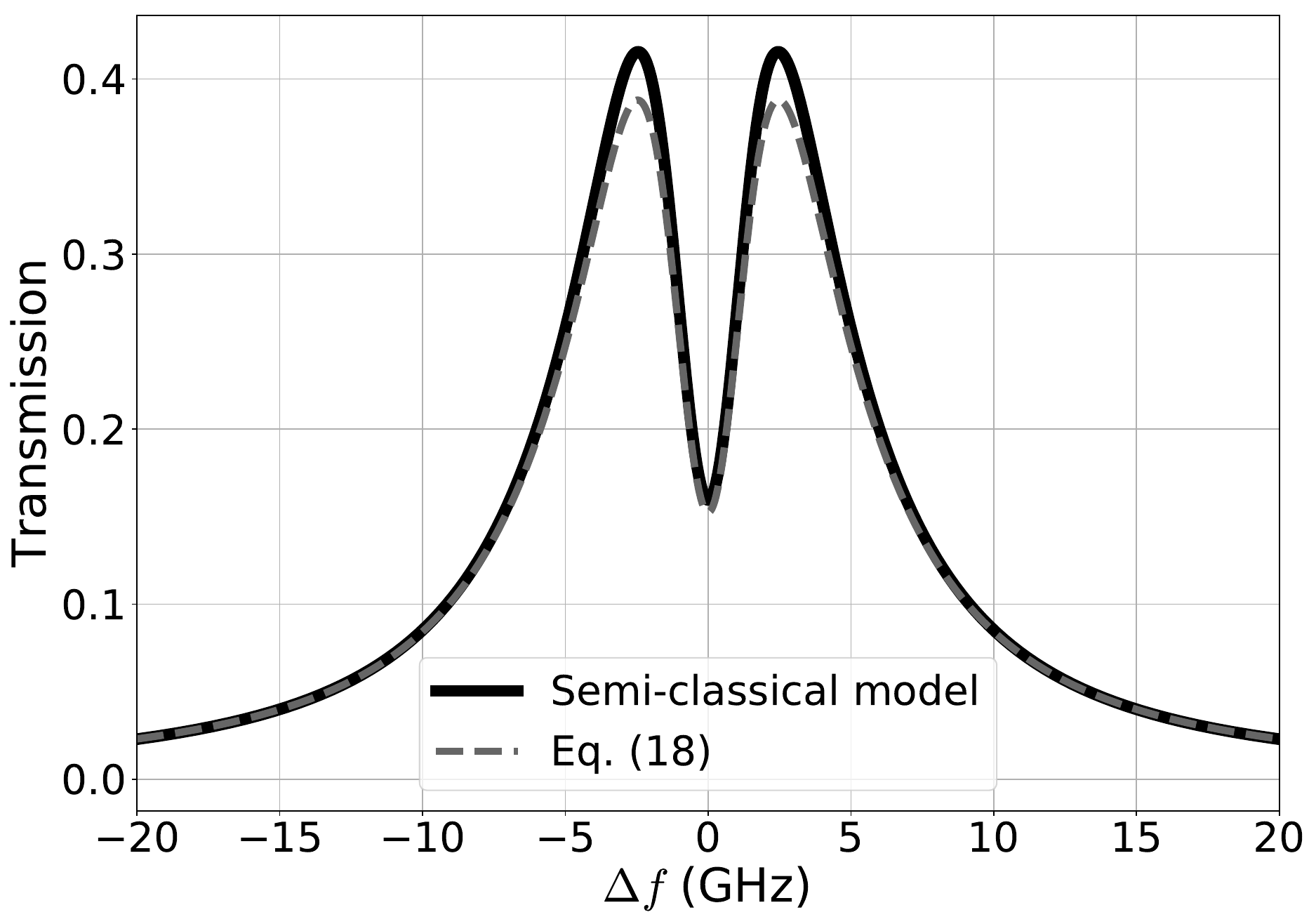}
\par\end{centering}
\caption{Comparison of the semi-classical model of Eq.~(\ref{Ch2-eq-transmission-electric-field-FP-1})
to the exact classical model of the lossy Fabry-Pérot cavity in Eq.~(\ref{Ch2-eq-intensity-fp-cavity-1})
for realistic parameters. The deviation between the dashed and solid
line is because in the semi-classical model only the first order effect
of absorption is taken into account.\label{Ch2-fig-compare-models-transmissionmodels-1}}
\end{figure}

To derive Eq.~(\ref{Ch2-eq-transmission-electric-field-FP-1}) from
Eq.~(\ref{Ch2-eq-transmissionsum-2}), we switch to transmission
(intensity) instead of the transmission amplitude (electric field).
This has the advantage that the imaginary parts disappear and we get
a better understanding of each term in the expansion. Using $1-R=t^{2}=1-r^{2}$,
we obtain from Eq.~(\ref{Ch2-eq-transmissionsum-2})

\begin{equation}
T_{tot}=\frac{\left(1-R\right)^{2}\exp(-2z)}{1+R^{2}\exp(-4z)-2R\exp\left(-2z\right)\cos\left(-2x_{1}+\phi\right)},\label{Ch2-eq-intensity-fp-cavity-1}
\end{equation}
with $z=a_{0}+a_{QD}\frac{1}{1+\left(\Delta^{'}\right)^{2}}$ and
$x_{1}=a_{QD}\frac{\Delta^{'}}{1+\left(\Delta^{'}\right)^{2}}$. Now
we use the following approximations: first, we consider small phase
changes $\phi\ll1$. This, in combination with $a_{QD}\ll1$, allows
us to approximate the cosine term as $\cos\left(-2x_{1}+\phi\right)\approx1-\frac{\left(-2x_{1}+\phi\right)^{2}}{2}.$
Trying to put the equation in a Lorentzian form gives

\begin{equation}
T_{tot}\approx\frac{1}{1+p_{0}+\left(\frac{-2x_{1}+\phi}{p_{1}}\right)^{2}},\label{Ch2-eq-intensity-fp-1-1}
\end{equation}
where $p_{1}=$$\frac{1-R}{\sqrt{R}}$ is related to the finesse of
an ideal Fabry-Pérot cavity ${F}=\frac{\pi\sqrt{R}}{1-R}$ and $p_{0}$
contains a contribution of loss due to the cavity and the QD. We neglect
$x_{1}^{2}$ in Eq.~(\ref{Ch2-eq-intensity-fp-1-1}) and find 
\begin{equation}
T_{tot}\approx\frac{1}{1+p_{0}+\left(\frac{\phi}{p_{1}}\right)^{2}-4\frac{x_{1}\phi}{p_{1}^{2}}}.\label{Ch2-eq-intensity-fp-approx-1-1}
\end{equation}
After Taylor expanding $p_{0}$ up to second order in $z$ we simplify
the analysis by splitting both loss terms and write $p_{0}=p_{c}+p_{QD}$
with

\begin{equation}
p_{c}=2a_{0}\left(\frac{1+R}{1-R}\right),\label{Ch2-eq-pcav-1}
\end{equation}

\begin{equation}
p_{QD}=2\frac{1}{1+\left(\Delta^{'}\right)^{2}}\left(a_{QD}+a_{QD}^{2}\right)\left(\frac{1+R}{1-R}\right).\label{Ch2-eq-pQD-1}
\end{equation}
For the cavity, we take $p_{c}$ up to first order in $a_{0}$ and
$p_{QD}$ up to second order in $a_{QD}$. This choice is made to
enable agreement with Eq.~(\ref{Ch2-eq-transmission-electric-field-FP}).
With this we can write Eq.~(\ref{Ch2-eq-intensity-fp-approx-1-1})
as

\begin{equation}
T_{tot}\approx\frac{1}{1+p_{c}}\frac{1}{1+\frac{p_{QD}}{1+p_{c}}+\frac{\phi^{2}}{p_{1}^{2}(1+p_{c})}-4\frac{x_{1}\phi}{p_{1}^{2}\left(1+p_{c}\right)}}.\label{Ch2-eq-intensity-fp-approx2-1}
\end{equation}
With the substitutions

\begin{equation}
\kappa=\frac{c(1-R)}{2\pi nL\sqrt{R}}\sqrt{1+p_{c}},\label{Ch2-eq-fp-kappa-1}
\end{equation}

\begin{equation}
\Delta=\frac{f-f_{c}}{\kappa}=\frac{1}{2\pi}\phi{F},\,\mathrm{and}\label{Ch2-eq-fp-Delta-1}
\end{equation}

\begin{equation}
C=a_{QD}\frac{\sqrt{R}}{1-R}\frac{1}{\sqrt{1+p_{c}}}\label{Ch2-eq-fp-C-1}
\end{equation}
we find for the total transmission

\begin{equation}
T_{tot}\approx\frac{1}{1+p_{c}}\frac{1}{1+4\Delta^{2}-8C\frac{\Delta\Delta^{'}}{1+\left(\Delta^{'}\right)^{2}}+\frac{2C}{1+\left(\Delta^{'}\right)^{2}}\left(2+2C\right)},\label{Ch2-eq-transmission-FP-Semiclass-1}
\end{equation}
where $\frac{p_{QD}}{1+p_{c}}\sim\frac{2C}{1+\left(\Delta^{'}\right)^{2}}\left(2+2C\right)$
assuming that $R\sim1$. Now we go back to the complex transmission
amplitude $t_{tot}=\sqrt{T_{tot}}$ of Eq.~(\ref{Ch2-eq-transmission-FP-Semiclass-1})
and arrive at Eq.~(\ref{Ch2-eq-transmission-electric-field-FP-1}).
In order to confirm that the above approximations are valid we compare
Eq.~(\ref{Ch2-eq-intensity-fp-cavity-1}) to the semi-classical model
of Eq.~(\ref{Ch2-eq-transmission-electric-field-FP-1}) in Fig.~\ref{Ch2-fig-compare-models-transmissionmodels-1}
for a micro-cavity with center wavelength $\lambda=930\,$$nm$, $n=2$,
$R=0.95$, $a_{0}=0.01$, $a_{QD}=0.03$, and $L=0.1\,$$\mu$m. We
see that both models agree very well, suggesting that our approximations
are valid. The slight deviations in the peak height is due to the
assumption that the cavity absorption $a_{0}$ is treated as a first
order effect in the semi-classical model.

\vspace{0.1cm}

\section{Comparison between the extended semi-classical and the quantum master
model}

\begin{figure}[H]
\includegraphics[width=1\columnwidth]{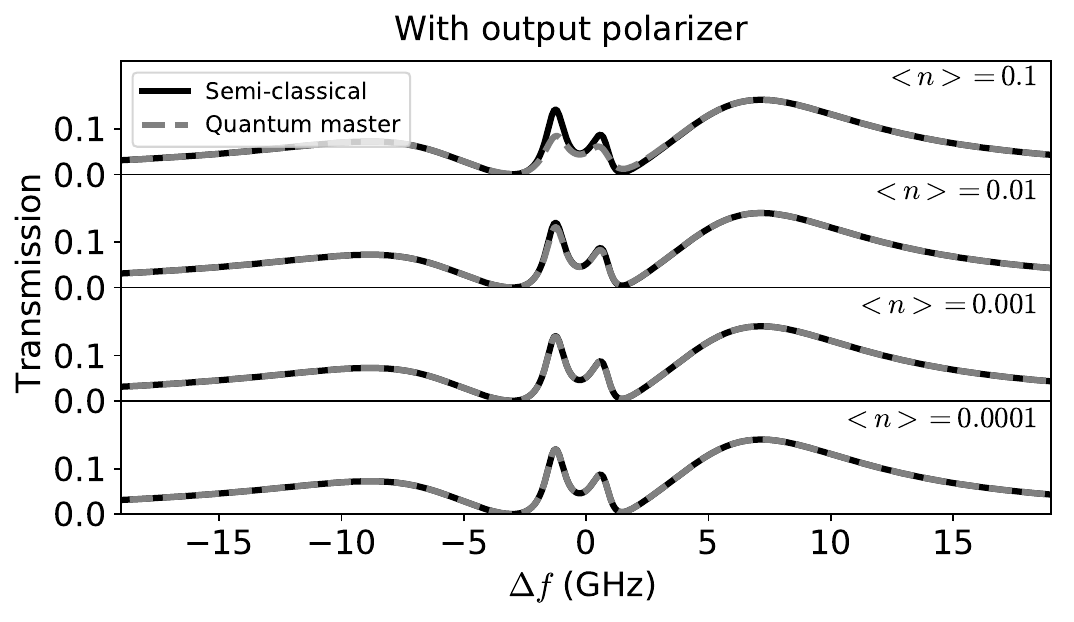}

\includegraphics[width=1\columnwidth]{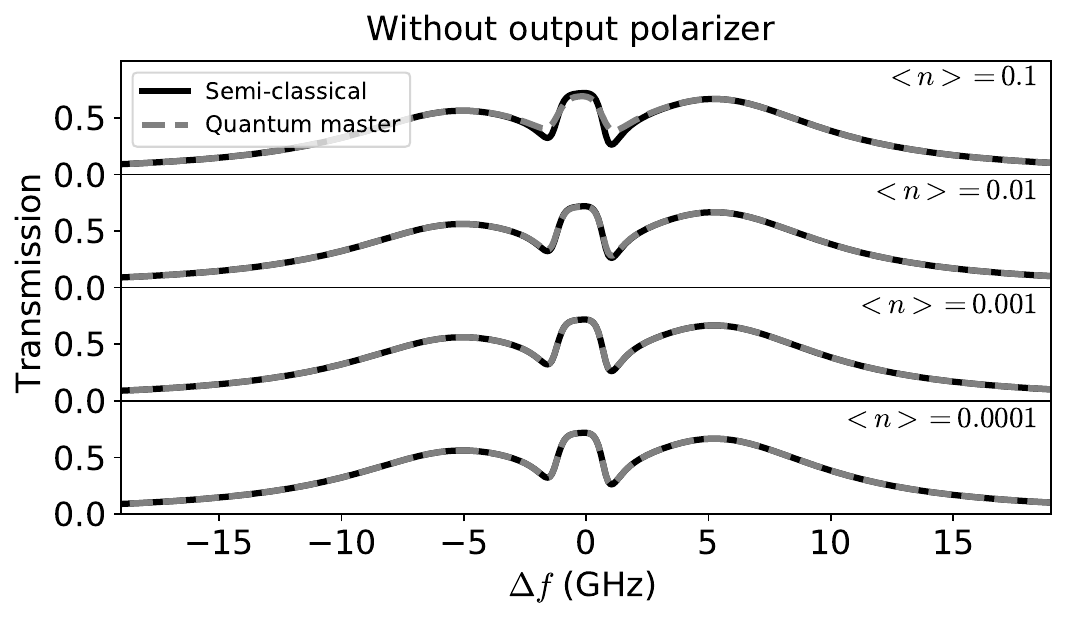}

\caption{Transmission spectra for the experimental case discussed in Figs.
\ref{Ch2-Fig3:crosspolplots} and \ref{Ch2-Fig4:Singlephotonrate}
of a neutral quantum dot with fine-structure splitting for the optimal
polarization condition (with and without output polarizer) at different
input power. Only at high input power ($\langle n\rangle=0.1$), our
extended semi-classical model deviates a bit from the quantum master
equation simulations.\label{fig6:transmissionandreflection}}
\end{figure}

Here we investigate the limit of our semi-classical model by increasing
the power to higher mean photon numbers. The mean photon number is
changed for the optimal polarization condition (Fig.~(3)) and the
condition where the output polarizer is removed. We see in Fig.~(\ref{fig6:transmissionandreflection})
that our results are similar up to $\langle n\rangle=0.01,$ for higher
power the results deviate due to saturation of the QD transitions.

\end{document}